**Angle-resolved photoemission study of USb$_2$: the 5f band structure**

E. Guziewicz[1,2], T. Durakiewicz[1,3], M. T. Butterfield[1], C. G. Olson[4]

J. J. Joyce[1], A. J. Arko[1], J. L. Sarrao[1], D. P. Moore[1], L. Morales[1]

[1]*Los Alamos National Laboratory, Los Alamos, New Mexico, 87545, USA*

[2]*Institute of Physics, Polish Academy of Sciences, 02-668 Warszawa, Poland*

[3]*Institute of Physics, Maria Skłodowska-Curie University, 20-031 Lublin, Poland*

[4]*Ames Laboratory, Iowa State University, Ames IA, USA*

ABSTRACT

Single crystal antiferromagnetic USb$_2$ was studied at 15K by angle-resolved photoemission with an overall energy resolution of 24 meV. The measurements unambiguously show the dispersion of extremely narrow bands situated near the Fermi level. The peak at the Fermi level represents the narrowest feature observed in 5f-electron photoemission to date. The natural linewidth of the feature just below the Fermi level is not greater than 10 meV. Normal emission data indicate a three dimensional aspect to the electronic structure of this layered material.

## 1. Introduction

The unusual properties of heavy fermion compounds have attracted considerable attention during the last two decades and several theoretical approaches have been developed in order to explain the large effective mass of the conduction electrons. It is well established that it is the f-electrons which are responsible for an effective mass enhancement and thus features that are specific to heavy fermion materials. The heavy fermion and antiferromagnetic ground states are both singlet states and similar in that the f-moment is compensated ostensibly by either a screening cloud of conduction electrons (heavy fermion) or an f-moment of opposite spin on an adjacent site (antiferromagnet). Photoemission techniques, which are capable of providing detailed information regarding the binding energy of the 5f electron band, as well as the dispersion and hybridization with the conduction band, are an especially valuable tool for evaluation of the various theoretical models.

U and Ce heavy fermion compounds exhibit similar bulk properties (magnetic susceptibility, resistivity and indications of enhanced mass), thus one might expect similarities in their electronic structures and similar theoretical models to be capable of explaining their heavy fermion behavior. The Single Impurity Model (SIM) [1 - 4] and the Periodic Anderson Model (PAM) [5, 6] have been the basic computational approaches, although many other models like the charge polaron model [7] or the two-electron band model [8] also address properties beyond one-electron models.

In the SIM model, f-electrons are treated as completely localized impurities in the sea of conduction electrons. This model assumes only slight hybridization with ligand conduction bands, which results in a non-dispersive f-levels. The second prediction of the model is that the PES f-electron weight scales with characteristic temperature. The shortcomings of this treatment of the PES data for correlated electron systems have been well documented [9 - 12].

The PAM model takes into account the coherent nature of electrons, thus, it may better describe the strong correlation of electrons in lattice systems. The complex nature of PAM calculations requires the application of some generalizations such as infinite dimensions, but initial results give certain general conclusions such as smaller temperature dependence of the f-band and its hybridization with conduction states. The PAM models are still generic and unable to calculate a specific strongly correlated system.

Though there are no specific calculations within the PAM for $USb_2$ (or for any element or compound for that matter), general predictions of the models can be assessed by experimental studies. To zero order, the photoemission results require a model which captures the periodicity of the lattice as well as the strong electron-electron interactions.

In principle, the distinction between localized and band-like behavior of f-electrons should be easily observable in PES experiments. However, the expected f-band dispersion is small, which makes the problem challenging to solve in practice. Any reasonable attempt to experimentally evaluate the 5f dispersion and weight needs to exhibit both very good energy and momentum resolution, and needs to take place on high-quality single crystals. The problem is observed in older PES data, where polycrystals measured at resonance (e.g. at photon energy over 100 eV) showed only a broad featureless structure pinned at Fermi edge, the so-called actinide triangle [13].

In the current study of $USb_2$ we have found a narrow feature near Fermi level which clearly exhibits dispersion. Dispersion was also observed in our normal emission photoemission data, giving evidence that $USb_2$, a layered compound, has some 3D character. The f-electron dispersion in the PES limits candidate models to those models which can accommodate periodicity with the lattice for a description of the electronic structure. The narrow feature at the Fermi level presented below is a true band feature with a natural linewidth less than the total dispersion of the feature in reciprocal space. Within this framework, the feature near $E_F$ in $USb_2$ is a true band state but renormalized to such an extent that the dispersion and natural linewidth are at least two orders of magnitude smaller than that in free electron models.

## 2. Experimental

We present ARPES data of USb$_2$ single crystals taken at low photon energies (20-60 eV) with an overall energy resolution between 24 meV (for hν= 34 eV) and 49 meV (for hν = 60 eV). The USb$_2$ crystals were prepared by the flux growth method and the PES studies were performed at the Synchrotron Radiation Center in Stoughton, Wisconsin.

We used a Plane Grating Monochromator (PGM) and estimate that the shifts of the spectral features near the Fermi edge, resulting from the PGM instrument function, would be of the order of 1meV. All measurements presented were taken at a constant temperature of 15K and thus standard temperature dependent effects may be neglected. The binding energy was referenced with a Pt Fermi level. The intensity was normalized to the mesh current at each data point to account for synchrotron beam decay and, after that, to the background intensity from higher order light giving rise to secondary emission above E$_F$ (as a means of normalize between different angles).

USb$_2$ is an antiferromagnet below 200K with a tetragonal layered Cu$_2$Sb-type structure (a = 4.270Å, c = 8.784 Å). This kind of structure allows cleaving with very little surface damage. Samples were cleaved under ultra-high vacuum (UHV) conditions to give smooth and flat surfaces for angle-resolved photoemission studies. Before introduction into the UHV photoemission chamber (p $\cong$ 10$^{-11}$Torr) the orientation of the sample was determined by X-ray diffraction. The high quality of the sample was confirmed by Laue patterns and by an extremely sharp photoemission peak near the Fermi level. PES spectra were measured using an angle-resolved analyzer with ±1$^0$ acceptance angle. The momentum resolution at hν = 30 eV is about 0.09 Å$^{-1}$, which is between 6.1% and 12.5% of the USb$_2$ Brillouin zone depending on the direction of investigation. The full width at half maximum (FWHM) of this peak in normal emission PES spectra taken at 34 eV photon energy and at 15K (Fig.1) is about 24 meV and increases to 49 meV for hν = 60 eV. The increase of FWHM between photon energy 34eV and 60 eV is the combined result of 1) reduced momentum resolution, 2) larger lifetime broadening that appears at higher photon energy 3) reduced electron analyzer performance at greater magnification values ($E_k/E_p$). Previous $USb_2$ data were taken with an energy resolution of 45 meV [14], so the apparent width of the photoemission structure at the Fermi edge was much greater, and dispersion relations and binding energies not as obvious. One of the significant improvements in this work is energy resolution near the value of the observed dispersion. The sharp near Fermi edge peak observed in the experiment is the sharpest photoemission feature found in uranium compounds up to date [15-20].

The nature of the sharp peak near Fermi level is shown in Fig.1. Several details are important; 1) The dotted lines are the Pt Fermi level and in all cases the peak in $USb_2$ is below $E_F$, 2) The peaks are very narrow but clearly show variation in binding energy as function of photon energy, 3) At the lowest energy (34 eV), the natural linewidth is conservatively estimated to be less than 10 meV when instrumental resolution is removed (and in reality may very possibly be much smaller than 10 meV). The dashed lines are the fitted data to the open circles after a Shirley background has been removed. The traditional Fermi liquid characteristics exhibit a E2 dependence of the linewidth. One may construct arguments for linewidths in excess of E2 arising from electron-electron, electron-phonon and electron-impurity interactions [21-22] and still maintain a Fermi liquid interpretation. However, the first peak linewidth being narrower than E2 would seem to place the interpretation of the electronic structure for USb2 outside the Fermi liquid regime.

## 3. Results and discussion

The $USb_2$ photoemission spectra presented in Fig. 2, 3 and 4 were taken for lower photon energies (34 eV, 43 eV and 60 eV) where good angular resolution and small lifetime broadening allow observation of dispersion in the U5f bands. This is a suitable photon energy range to achieve an adequate compromise between the high U5f photoionization cross-section while still retaining substantial momentum resolution.

Spectra taken at each photon energy were normalized to the intensity above the Fermi edge arising from higher order secondaries. The [001] surfaces were oriented by use of a Laue X-ray camera. The electron energy analyzer was varied between $\theta = 0^0$ and $\theta = 6\text{-}10^0$, which corresponds to the $\Gamma$-X direction in the Brillouin zone.

In the photon energy range of 34-60 eV the photoionization cross-section of U5f increases dramatically with hν and for 60 eV is about twice as high as for 34 eV [23]. In turn, the U6d and the Sb5p (the only shells with cross-sections comparable to the U5f) cross-sections show the opposite photon energy dependence and for hν = 60 eV is approximately three times lower than for hν = 34 eV. The main feature of all of the spectra presented in Fig. 2a, 3 and 4 is the sharp but dispersive peak near the Fermi edge (see Fig. 2b). The structure labeled B situated between 300 meV and 600 meV appears to grow with increasing photon energy up to 60 eV in normal emission PES spectra (Fig.4), unlike the structure at the Fermi edge (A). Using cross-sections argument we relate the feature B mainly to the U5f emission. However, because the structure B is broader than normally ascribed to 5f peaks, we propose that it has a mixed conduction band-5f origin.

By comparison to similar materials containing Sb and f electrons, we do not expect any substantial admixture of the Sb5p state near the Fermi edge. Currently there are no theoretical calculations of the $USb_2$ electronic structure, so we base our assumption on the theoretical and experimental results of USb, CeSb and $CeSb_2$ [24-26]. The comparison of theoretical calculations and photoemission data on USb shows that the U5f state has itinerant rather than localized character. In the itinerant model the Sb5p bands are totally occupied, dispersive, and located 1-4 eV below the Fermi level. There is only a small overlap between the Sb5p and U6d bands. This overlap has little influence on the electronic structure of USb near the Fermi level. We assume similar characteristics for $USb_2$. Although the crystal structure is different, in $USb_2$ the occupied U5f state is also close to the Fermi level and consequently located above the Sb5p band. The interaction between the U5f and the Sb5p electrons results in pushing the Sb5p state towards the higher binding energy [25]. Therefore we would expect that within 0-1 eV below the Fermi level the electronic band structure of USb$_2$ is dominated by U6d and U5f states.

The normal emission PES spectra measured for hν = 34 eV (Fig.2a) shows one sharp feature situated near the Fermi edge. The rest of the spectrum remains completely flat, which is an evidence of a very clean, high quality sample surface resulting in very well defined incidence and emission angles. The peak near the Fermi edge, which was interpreted so far as the U5f-conduction band hybridized narrow band, changes its position from 37 meV for θ = 0$^0$ to 23 meV for θ = 6$^0$ and 7$^0$ giving evidence of a 14 meV dispersion (Fig. 2b). For θ = 5$^0$ the FWHM of this peak starts growing from 24 meV for lower angles, up to 48 meV for θ = 7$^0$. The given FWHM values include the instrumental resolution and thus the natural line width is extremely sharp (as demonstrated in the detailed analysis of Fig.1). When fitting the data we use a Gaussian function for experimental resolution and a Lorentzian function for the natural line width. For θ = 8$^0$ and 9$^0$ we see two structures in the Fermi level region. It appears that the increase in width away from normal is a result of two states with different dispersions.

A large dispersion (about 270 meV) is shown in the structure marked as B in Fig. 2b, which appears at 470 meV below the Fermi level for θ = 4$^0$. The intensity of this structure grows gradually up to θ = 9$^0$, whereas the binding energy shifts downwards and for θ = 9$^0$ it is situated at 200 meV.

The PES spectra taken for hν = 43 eV (Fig.3) also shows dispersion of peak A, which changes the energy position from 34 meV (θ = 0$^0$) to 48 meV (θ = 6$^0$). We can not see two peaks for θ = 6$^0$, but the FWHM is almost twice of that for θ = 5$^0$, which suggests an additional contribution for higher angles. The structure B appears for θ = 2$^0$ at a binding energy of 410 meV and changes its binding energy position to 234 meV for θ = 6$^0$.

Photoemission data for hν = 60 eV (Fig.4) show both A and B structures even in the normal emission spectrum. The dispersion of peak A in this case is 14 meV, the A position for θ = 0$^0$ is 67 meV and for θ = 10$^0$ is 81 meV below E$_F$. The structure B is situated at higher binding energies than for photon energy 34 eV and 43 eV. Its energy position varies from 575 meV ($\theta = 0^0$) to 337 meV ($\theta = 10^0$).

The position and intensity of peak B is different for the three photon energies investigated in Figures 2a, 3 and 4. This partially derives from the fact that we probe different parts of the Brillouin zone. The set of spectra shown in Fig. 2 are taken near the Z point in the Brillouin zone, whereas those in Fig.3 and 4 are both taken close to the Γ point. In Fig.4 the peak is more pronounced that in Fig.3 and also appears at a higher binding energy, even though at both these photon energies we probe the vicinity of the Γ point. We believe that this is a matrix element effect which has stronger influence on the spectra taken for hν = 43 eV. In the case of hν = 60 eV the final states are more free-electron like and hence the photoemission spectra are less influenced by matrix element effects.

Photoemission is a surface-sensitive experiment. Therefore there always exists a question as to whether features observed in the valence band derive from the bulk electronic structure or the surface electronic structure, which may differ considerably. We investigated the surface-bulk problem by means of a controlled surface termination experiment (see Fig. 5). We cleaved a $USb_2$ sample and exposed it to up to $1 \times 10^{-8}$ Torr of argon for 20 seconds and observed changes in the valence band by taking EDCs at a photon energy of 22eV. The effect was mostly attenuation of the valence band features. After exposure for 100 seconds (p=$10^{-8}$ Torr of Ar) the photoemission from the valence band has almost disappeared, and the Fermi level peak disappears at the same rate. We carried out the surface termination and inert absorbate experiment for multiple angles to see if the changes are fundamental or are a result of k scattering. We observed that the changes are similar for different angles i.e. in the investigated part of the Brillouin zone the strong Fermi level peak is not surface related. We noticed that at 22 eV, photon exposure was causing the sample to revert to the non-absorbate surface. Therefore we warmed the sample to 40K. We observed that the Ar3p peak disappeared and the valence band returned to an attenuated version of the baseline spectra. The controlled absorbate experiment showed that the shape of the valence band photoemission spectra near the Fermi edge remains the same with respect to the other valence band features and one observes the predictable exponential decay of the entire valence band with PES mean free path as Ar is covering the surface.

Normal emission spectra presented in Fig. 6 provide additional evidence that the near $E_F$ peak derives from bulk crystal states. The data were taken for photon energies ranging from 17.54 eV to 34 eV. A dispersion of around 10 meV of the sharp near-$E_F$ peak may be seen. Dispersion perpendicular to the surface is evidence that the very sharp peak near the Fermi edge is not a surface state. This also means that $USb_2$ is not of purely 2D electronic structure but has a 3D character which couples weakly to the in-plane features, and thus requires treatment as a 3D material in reciprocal space. This is in contrast to the Fermi surface proposed in [27-28] which is decidedly 2D.

Measurements near the U5d → U5f absorption edge (Fig.5) were done at 15K. The main resonance in uranium compounds is split into two because of the large spin-orbit splitting of the U5d shell. Consequently, there is no clean anti-resonance in uranium compounds, but just a minimum of the resonances from each of the cores. The PES spectra taken near the maximum of the resonance (hν=108 eV) shows the U5f enhancement mainly in the binding energy range between 300 meV and the Fermi edge. The spectra measured near the valley between the 3/2 and 5/2 resonances (hν = 102 eV) shows the main photoemission structure to be between 300 meV and 500 meV, but we can also notice a smaller structure near the Fermi edge. These results show that the B structure, which is also observed in PES spectra taken at lower photon energies, consists of hybridized 5f and conduction band electrons. The peak A is primarily of 5f origin but contains a non negligible contribution from the conduction electrons.

Resonant photoemission measurements confirm the conclusion about the 5f hybridization. The small intensity difference between on- and off-resonance spectra is indicative of 5f-conduction band hybridization, but also that the valley at 102 eV is not a true antiresonance, but just a minimum between the two main resonances at 108 eV and 98 eV. Photoemission spectra and calculations for uranium compounds [29] show larger transition probabilities from the almost pure U5f states at $E_F$ than from hybridized U5f bands.

One should remember that resonant photoemission data have qualitative rather than quantitative character. This is due to the different kinds of Auger processes that occur in rare earth and actinide materials [30] (because of the vicinity of partially occupied f and d shells) which give rise to additional decay channels. In uranium-bearing materials there are unique Auger processes based on 5f→$E_F$ transition [31], so the resonant Auger decay can be comparable in strength to direct resonant photoemission. Taking into account all of the processes mentioned above it is clear that in the case of uranium compounds the resonant photoemission experiment provides qualitative information about conduction band-5f hybridization. Also, the reduced momentum resolution at photon energies over 100 eV precludes observation of the subtle U5f changes. However, the resonant photoemission experiment confirms that the A and B photoemission peaks have conduction band and 5f origin, giving evidence of conduction band-5f hybridization.

Our results show that the near $E_F$ 5f photoemission features of $USb_2$ behave in a way similar to the 4f features of Ce compounds. For example, the ARPES spectra of $CeBe_{13}$ [10], $CeSb_2$ [8] and $CePt_{2.2}$ [32] show a sharp 4f peak near the Fermi edge. For $CeBe_{13}$ the evidence for dispersion of the 4f band was found for two directions of the Brillouin zone. In the case of $CeSb_2$ the dispersion, if present, is expected not to be larger than 10 meV. However, both the $4f_{5/2}$ and $4f_{7/2}$ bands are strongly momentum dependent, suggesting band-like behavior.

One could expect differences between $USb_2$ and $CeSb_2$ in the location of Sb5p states as is the case of USb and CeSb [24-26]. In CeSb the Sb5p electrons are located closer to the valence band edge because of their different energy position relative to the bare 4f state. However, this assumption needs to be verified by the comparison of photoemission results with theoretical calculations. The near Fermi level part of the electronic structure of $USb_2$ and $CeSb_2$, as seen in photoemission experiment, is indeed similar. Ce and U heavy fermion compounds display similar bulk properties and similar band structure as well. Therefore it is reasonable to assume that they might be described within a similar theoretical framework. Our experimental results are consistent with the Periodic Anderson Model. The evidence of hybridization between conduction band and U5f electron states in $USb_2$ presented above supports this assumption.

## 4. Conclusions

Our angle-resolved photoemission studies of $USb_2$ crystals, taken with an energy resolution of 24 meV, unambiguously show the dispersion of the U5f – conduction band hybridized bands. The contribution of the 5f electron density of states near the valence band edge and in the binding energy region of 300-600 meV was confirmed by both photoionization cross-section dependencies and resonant photoemission. The dispersion of the band (A) closest to the Fermi edge was found to be 14 meV, whereas the broader structure situated in the binding energy range 300-600 meV shows dispersion between 200 and 260 meV. There is also around 10 meV of dispersion in the normal emission data which is an indication that $USb_2$ has some 3D character. The in-plane bonding appears dominant over the c-axis bonding by virtue of the larger dispersions observed in peak B in-plane.

The results presented show substantive similarities between Ce and U compounds [9] and suggest that a similar theoretical framework might be used to describe these two kinds of correlated f-electron systems. The band-like behavior of the U5f electrons and 6d-5f binding energy sequence are in qualitative agreement with a Periodic Model, including PAM, but other periodic models must be considered.

**Acknowledgment**
This work was supported by the US Department of Energy, Office of Science, Division of Materials Science and Engineering and under contract W-7405-ENG-82. This work is based upon research conducted at the Synchrotron Radiation Center, University of Wisconsin-Madison, which is supported by the NSF under Award # DMR-0084402.
.

**Figure captions**

Fig.1. Comparison of the Fermi edge of Platinum with near – $E_F$ features of $USb_2$. The dashed lines are the fitted data to the open circles (after a Shirley background has been removed). The FWHM values are defined mainly by instrumental resolution. The natural line width is extremely narrow; the width calculated for hν = 34 eV is below 10 meV.

Fig.2. a) High-resolution angle-resolved photoemission spectra of $USb_2$ within 800 meV of $E_F$ taken at hν = 34 eV,
b) The same valence band spectra of $USb_2$ within the first 0.08 eV of $E_F$.

Fig.3. High-resolution angle-resolved photoemission spectra of $USb_2$ within 800 meV of $E_F$ taken at hν = 43 eV.

Fig.4. High-resolution angle-resolved photoemission spectra of $USb_2$ within 800 meV of $E_F$ taken at hν = 60 eV.

Fig.5. Normal emission spectra of $USb_2$ taken in a controlled surface termination experiment.

Fig.6. Normal emission spectra of $USb_2$ within 70 meV of $E_F$.

Fig.7. Normal emission spectra taken at Fano resonance (108 eV) and antiresonance (102 eV) at 15K.

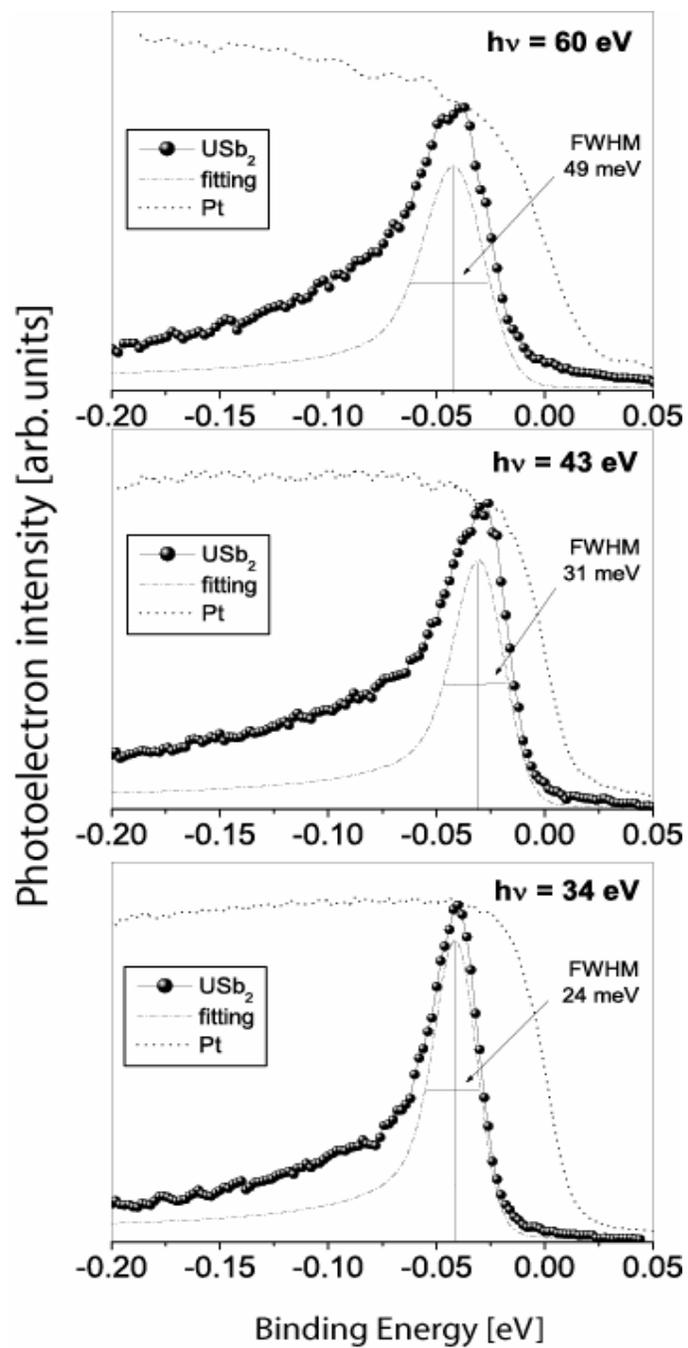

Fig. 1.

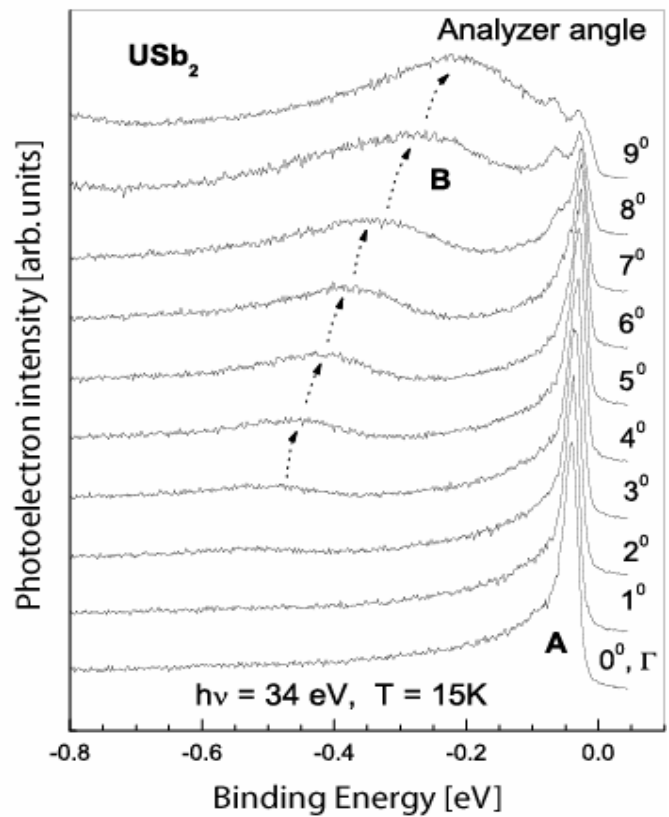

Fig. 2a

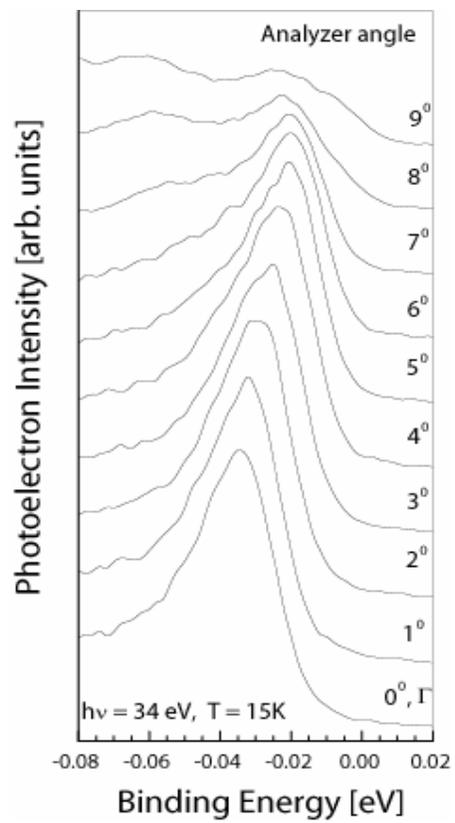

Fig. 2b

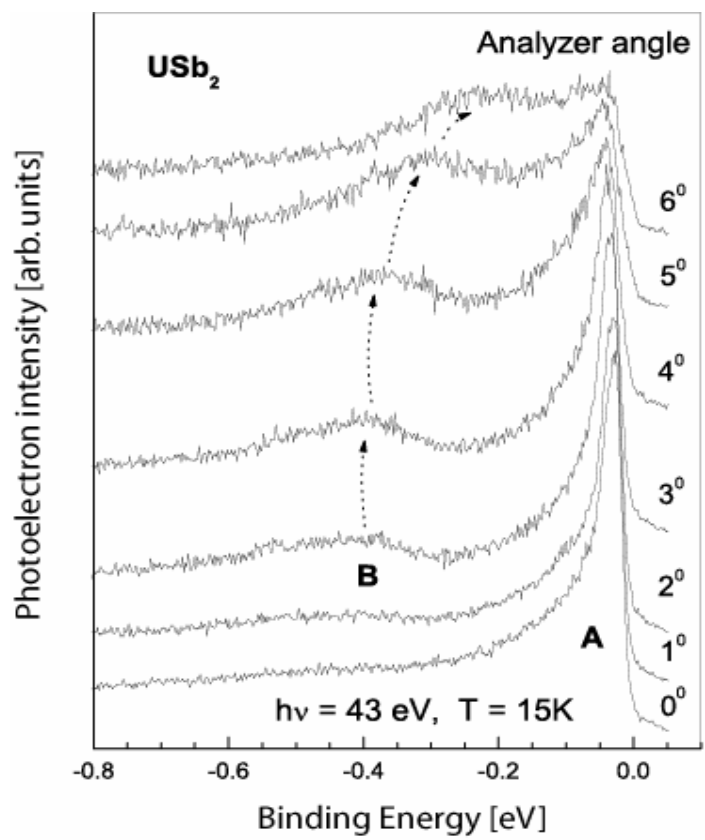

Fig. 3.

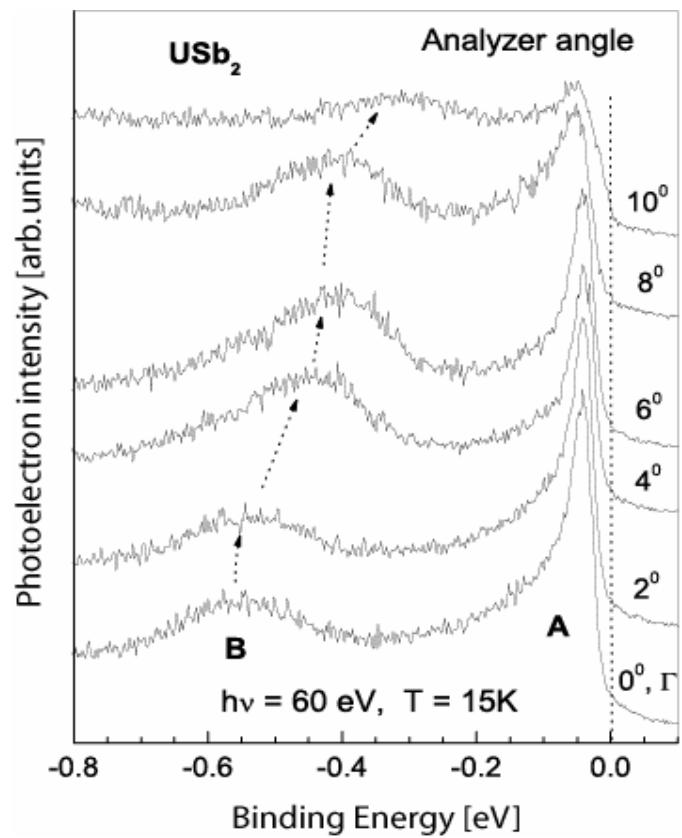

Fig. 4.

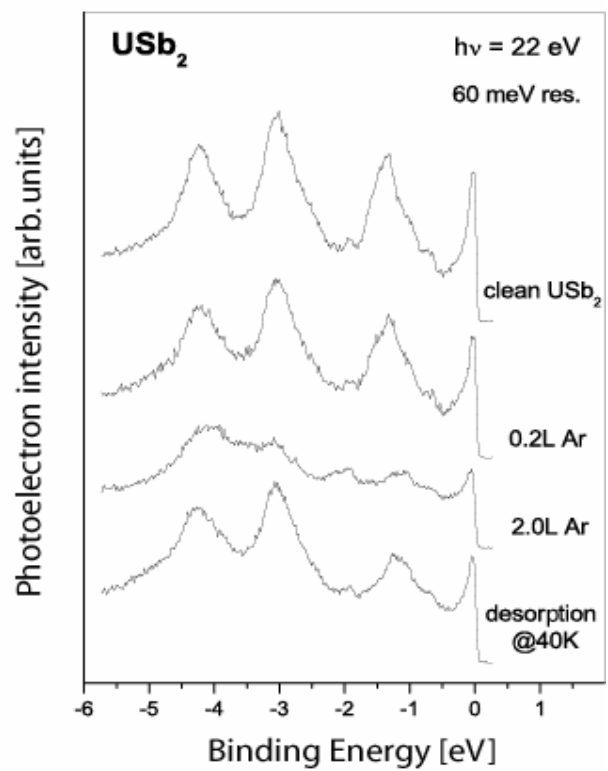

Fig. 5.

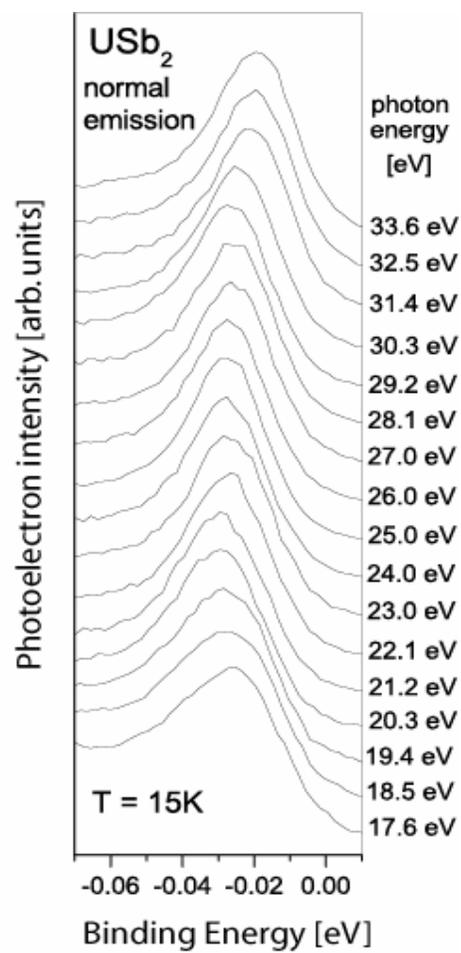

Fig. 6.

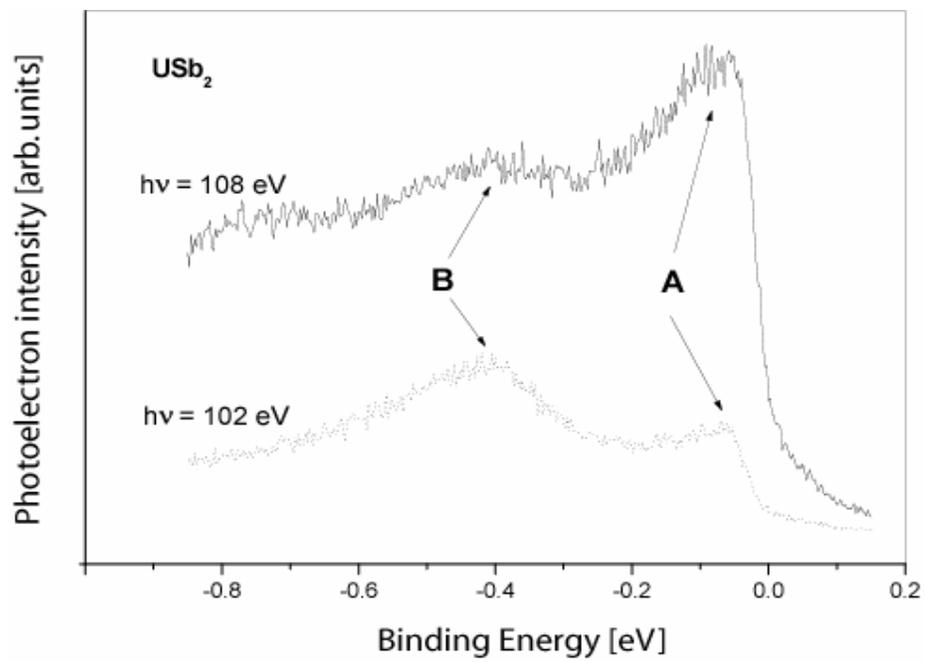

Fig. 7.